\documentstyle[amssymb,12pt,psfig,epsfig]{article}

\def\be{\begin{equation}}
\def\ee{\end{equation}}
\def\ba{\begin{eqnarray}}
\def\text#1{\mathrm{#1}}
\def\ea{\end{eqnarray}}

\newcommand{\Zz}{\ifmmode {\rm Z} \else ${\rm Z } $ \fi}

\begin{document}
\date{February 6, 2002}
\title{Precision Test of Quark Mass Textures: A Model Independent Approach}
\author{F. Caravaglios$^a$,  P.\ Roudeau$^b$,  A. Stocchi$^{b,c}$ \\
{\small $^a$ Dipartimento di Fisica, Universit\`{a} di Milano, {\it and \ }INFN\
sezione di Milano} \\
 {\small$^b$ Laboratoire de l'
Acc\'el\'erateur Lin\'eaire {\it IN2P3-CNRS et Universit\'e de Paris Sud }}
\\
\small $^c$ CERN, CH-1211 Geneva 23, Switzerland}
\maketitle

\begin{abstract}
Using a Monte Carlo method, we have directly extracted from the available 
measurements, the hierarchies among the different elements of the quark mass
matrices. To do that, we have first introduced a model independent 
parameterization for two generic class of models: those based on Abelian 
symmetries and those inspired by a $U(2)$ horizontal symmetry. So, matrix 
entries are proportional to  some $\epsilon ^{t}$, with $\epsilon<<1$  and
 the $t$'s are different free exponents
that we determine from the data through a statistically well defined 
procedure. We have found that the experimental data poorly constrain the 
Abelian scenarios.

Instead, in non Abelian scenarios, these $t$-exponents are strongly 
constrained by the present data.We have found that contrary to a {\it naive} 
$U(2)$ horizontal symmetry expectation, \ quark mass matrices turn out to be
not symmetric. Two solutions emerge: one with $M_{32}^{{\rm down}}\gg 
M_{23}^{{\rm down}}$ and $M_{21}^{{\rm up}}\gg M_{12}^{{\rm up}}$ ; and a 
second one with slight asymmetries only in the light quark sector, namely $
M_{21}^{{\rm up}}<M_{12}^{{\rm up}}$ and $M_{21}^{{\rm down}}>M_{12}^{{\rm %
down}}$ .
\end{abstract}

\vskip 0.5cm 
{IFUM-702-FT}

\vfill 
\eject

\section{Introduction}

Standard Model predictions have been tested so far with very high accuracy
(at the permill level), and the agreement with the experimental measurements
is very remarkable. However the Standard Model is unsatisfactory since all
fermion masses and mixings are free and unpredictable parameters. Quark
masses vary in a range of more than four order of magnitudes and the lack
for an explanation is deceiving. \newline
If the Higgs boson is a $SU(2)$ doublet, the four real components 
\begin{equation}
\Phi =\pmatrix{\phi_1+ \mathrm{i} \phi_2 \cr \phi_3+ \mathrm{i} \phi_4 }
\label{eq:1a}
\end{equation}
can be transformed by an $SO(4)\equiv SU(2)_{{\rm left}}\times SU(2)_{{\rm %
right}}$ symmetry. The unique $SU(2)$ invariant $\Phi ^{\dagger }\Phi =\phi
_{1}^{2}+\phi _{2}^{2}+\phi _{3}^{2}+\phi _{4}^{2}$ is also an $SO(4)$
invariant. The Higgs potential must be a $SO(4)$ invariant, and when the $%
SU(2)\times U(1)$ symmetry is broken into $U(1)_{em}$, the $SO(4)$ symmetry
is broken into $SU(2)_{V}=SU(2)_{{\rm left+right}}$. The Goldstone bosons of
the Higgs doublet transform as a triplet under the unbroken $SU(2)_{V}$ and
this directly implies a fundamental relation between the mixing angle of the
neutral gauge bosons and the charged/neutral boson masses \cite{peskin} 
\begin{equation}
1-\frac{M_{W}^{2}}{M_{Z}^{2}}=\sin ^{2}\theta _{W}.
\end{equation}
The remarkable experimental success\footnote{%
After the inclusion of some expected and calculable radiative corrections.
See \cite{cara} and references therein.} of this relation which has been
proved by the precision tests at $LEP$ and $SLD$, makes the above $SO(4)$
symmetry unique and unavoidable for any realistic description of the
electroweak symmetry breaking. The same success is far from being achieved
in the fermion sector, because there are too many models and symmetries
which roughly give acceptable predictions ({\it i.e.} in terms of order of
magnitudes). The large number of fermions and the lack of direct
measurements of all the mass matrix entries make particularly hard to
unambiguously select the underlying symmetries responsible for fermion mass
hierarchies: measuring the CKM matrix elements, one fixes only the product
of two unitary matrices, coming from the separate diagonalizations of the up
and down $3\times 3$ quark mass, and not both of them. \newline
To cope with these ambiguities, one usually assumes a fundamental symmetry
and chooses a breaking pattern. The mass matrices obtained are diagonalized.
Eigenvalues and eigenvectors are compared with the experimental data, {\i .e.%
} the CKM\ parameters and the quark masses. This procedure can be used to
rule out a model or a symmetry if theoretical predictions and experimental
data disagree. It is, of course, unable to prove that the assumed symmetry
is the one which is preferred by the data{\bf . } Ideally one could take a
completely model independent approach and prove that a symmetry is among the
best ones that can predict the correct mixings and masses spectrum. This
goal is probably unachievable, but a step in this direction can be done if
one accepts a certain amount of model dependence, selecting a large class of
models, and then extracting directly from experimental results the preferred
models within that class.

In the following, we explicitly discuss how to approach this goal,
introducing model independent parameterizations for the up and down quark
mass matrices. The needed number of free parameters \ will be in general
larger than the number of experimental constraints and this will be the main
limitation for a direct fit of parameter values\footnote{%
The minimum of the $\chi ^{2}$ would be not a single point but obscure and
intricate surfaces in the $n$-dimensional space of the $n$ parameters.}.
Instead of minimizing the $\chi ^{2}$ \ with respect to the full set of \
free parameters (which is too large to be manageable), we get rid of several
unknown parameters, treating them as real theoretical errors, through a
Monte Carlo procedure.

To exemplify the proposed method, we will focus on two different
parameterizations inspired by two classes of models: those based on Abelian
symmetries and those exploiting a horizontal $U(2)$ symmetry.

\section{The experimental constraints}

\label{sec:exptinput}

Values of the six quark masses and of the four CKM matrix element
parameters, used in the present analysis are given in Tables \ref{tab:massq}
and \ref{tab:ckm}.

\begin{table}[th]
\begin{center}
\begin{tabular}{|c|c|c|}
\hline
Quark flavour & Reference value & $\overline{m}_q(m_W)$ \\ 
& GeV/c$^2$ & GeV/c$^2$ \\ \hline
top & ${\rm M}^{pole}_t= 174.3 \pm 5.1$ & $175.3 \pm 5.1 \pm 0.2$ \\ 
bottom & $\overline{m}_b(\overline{m}_b)=4.19 \pm 0.05$ & $2.906 \pm 0.035
\pm 0.031$ \\ 
charm & $\overline{m}_c(\overline{m}_c)=1.304 \pm 0.027$ & $0.667 \pm 0.014
\pm 0.023$ \\ 
strange & $\overline{m}_s(m_{\tau})=0.125 \pm 0.030$ & $0.072 \pm 0.017 \pm
0.002$ \\ \hline
up and down quarks & Reference values & $\overline{m}_q(m_W)~{\rm MeV/c^2}$
\\ \hline
& $\frac{m_u}{m_d}=0.46 \pm 0.09$ & $\overline{m}_d(m_W)=3.7 \pm 0.9$ \\ 
& $Q^2=\frac{4m_s^2-(m_u+m_d)^2}{4(m_d^2-m_u^2)}=22.0 \pm 0.6$ & $\overline{m%
}_u(m_W)=1.7 \pm 0.5$ \\ \hline
\end{tabular}
\end{center}
\caption{{\it {Values of quark masses used in this analysis. The second
column gives the reference values from which they are evolved using the $%
\overline{{\rm MS}}$ scheme. For quarks heavier than $u$ and $d$ the last
column gives the values extrapolated at the scale of the W boson mass, the
first uncertainty corresponds to the uncertainty on the reference values and
the second uncertainty is due to the extrapolation. Uncertainties quoted for 
$u$ and $d$-quark flavours are only indicative as mass ratios are directly
determined.} }}
\label{tab:massq}
\end{table}

\begin{table}[ht!]
\begin{center}
\begin{tabular}{|c|c|}
\hline
CKM parameter & Reference value \\ \hline
$\lambda$ & $0.2237\pm 0.0033$ \\ 
$\left | {\rm V}_{cb} \right |$ & $(41.0 \pm 1.6)\times 10^{-3}$ \\ 
$\rho$ & $0.225 \pm 0.038$ \\ 
$\eta $ & $0.317\pm 0.041 $ \\ \hline
\end{tabular}
\end{center}
\caption{{\it {\ Values of the CKM matrix parameters used in the present
analysis \cite{ref:ciuch}.} }}
\label{tab:ckm}
\end{table}

\subsection{Quark mass determinations}

Values of quark masses span a large domain ranging from a few MeV/c$^2$, for 
$u$ and $d$ flavours, to more than 100 GeV/c$^2$ for the top quark. Current
quark masses, defined in the $\overline{{\rm MS}}$ scheme have been used and
their values have been given at the scale of the W mass. It can be noted
that the present analysis depends on quark mass ratios and that these
quantities are almost scale independent. The running of quark masses has been
evaluated following the work of \cite{ref:koide}. The value adopted for the
strong coupling constant, at the scale of the $\ifmmode {\rm Z} \else ${\rm %
Z } $\fi$ boson mass is $\alpha_s(m_Z)=0.1181 \pm 0.0020$ which corresponds
to $\Lambda^{(5)}=209.5^{+24.4}_{-22.6}$ MeV in the $\overline{{\rm MS}}$
scheme.

The top-quark mass has been measured from the direct reconstruction of its
decay products \cite{ref:top}. It corresponds to the pole mass value: 
\begin{equation}
{\rm M}^{pole}_t=(174.3 \pm 5.1)~{\rm GeV/c}^2.
\end{equation}
The expression relating $\overline{m}_t({\rm M}^{pole}_t)$ and ${\rm M}%
^{pole}_t$ can be found in \cite{ref:koide}.

Lattice determinations of the $b$ and $c$  
quark masses can be found in \cite{quarkmasses}.
The $b$-quark mass is obtained using QCD sum rules for the $\Upsilon$ masses
and electronic widths following the work of \cite{ref:volo}. Recent
evaluations and complete references can be found in \cite{ref:hoang}
 giving: 
\begin{equation}
\overline{m}_b(\overline{m}_b)=(4.17 \pm 0.05)~{\rm GeV/c}^2.
\end{equation}
A new analysis \cite{ref:kuhn} of the energy dependence of hadronic
production near the threshold for $b$-quark production gives a similar
value: 
\begin{equation}
\overline{m}_b(\overline{m}_b)=(4.209 \pm 0.050)~{\rm GeV/c}^2.
\end{equation}
In the following the value $\overline{m}_b(\overline{m}_b)=(4.19 \pm 0.05)~%
{\rm GeV/c}^2$ has been used.

The $c$-quark mass was also determined using QCD sum rules applied to the
charmonium system, for a recent review see \cite{ref:jamin} in which they
obtain: 
\begin{equation}
\overline{m}_c(\overline{m}_c)=(1.23 \pm 0.09)~{\rm GeV/c}^2.
\label{eq:jamin}
\end{equation}
The same analysis \cite{ref:kuhn} which was applied to the $b$-quark mass
determination was done in the charm threshold region and gives: 
\begin{equation}
\overline{m}_c(\overline{m}_c)=(1.304 \pm 0.027)~{\rm GeV/c}^2.
\end{equation}
This value is compatible with Equation (\ref{eq:jamin}) and is more precise;
it has been adopted in the following.

The value of the strange quark mass has been obtained \cite{ref:davier} \cite
{ref:korner} from analyses of $\tau$ decays into final states with an odd
number of kaons which give, respectively: 
\begin{equation}
\overline{m}_s(m_{\tau})=(120 \pm 11_{exp}\pm8_{|V_{us}|} \pm 19_{th})~{\rm %
MeV/c}^2  \label{eq:davier}
\end{equation}
and 
\begin{equation}
\overline{m}_s(m_{\tau})=(130 \pm 27_{exp}\pm 8_{|V_{us}|} \pm 9_{th})~{\rm %
MeV/c}^2.  \label{eq:kun}
\end{equation}
The different values quoted for systematic uncertainties of theoretical
origin depend on the order of the moments of the spectral function which
have been used in the two analyses. In the following the value $\overline{m}%
_s(m_{\tau})=(125 \pm 30)~{\rm MeV/c}^2$ has been used. It corresponds to a
central value of $159$ MeV/c$^2$ when evaluated at the scale of 1 GeV.

Finally values for light quark masses can be obtained from chiral
perturbation theory \cite{ref:leut}. Recent developments have been included 
\cite{ref:bijn} in the values adopted in the following: 
\begin{equation}
\frac{m_{u}}{m_{d}}=0.46\pm 0.09;~Q^{2}=\frac{4m_{s}^{2}-(m_{u}+m_{d})^{2}}{%
4(m_{d}^{2}-m_{u}^{2})}=22.0\pm 0.6  \label{eq:udr}
\end{equation}
From these values, the ratio between the masses of the $s$ and $d$ quarks is
obtained:$\frac{m_{s}}{m_{d}}=19.5\pm 1.2$. Using the determination of $%
\overline{m}_{s}(m_{W})$ given in Table \ref{tab:massq}, values are obtained
for light quark masses at the scale of the W mass. 

\subsection{CKM matrix parameters}

In the Wolfenstein parametrisation of the CKM matrix \cite{ref:wolf}, the
four parameters are designed, usually, as: $\lambda$, $A$, $\rho$ and $\eta$%
. In the following, in place of $A$, the quantity, $\left | {\rm V}_{cb}
\right |=A \lambda^2$ has been used instead to avoid correlations.

Updated determinations of their values, which are given in Table \ref
{tab:ckm}, can be found in \cite{ref:ciuch}. The values of $\rho$ and $\eta$
have been determined within the Standard Model framework using the
constraints on these parameters coming from ${\rm {B^{0}_{d}}}$ and ${\rm {%
B^0_s}}$ oscillations, $b \rightarrow u$ semileptonic decays and from CP
violation in K and B mesons \footnote{%
This last constraint, which was not available in \cite{ref:ciuch} has been
included in the present analysis.}.

\subsection{Procedure adopted to apply constraints}

To apply the constraints related to quark masses and CKM elements we have
defined the function 
\begin{equation}  \label{eq:chisquare}
\chi^2(O^{{\rm t}h})=\sum_{i}\left( \frac{O^{{\rm t}h}_i- O^{{\rm e}xp}_i}{%
\sigma^{{{\rm e}xp}}_i} \right)^2.
\end{equation}
where $O^{{\rm e}xp}_i\pm \sigma^{{\rm e}xp}_i$ are the experimental inputs
as shown in Tables 1 and 2. The function $\chi^2(O^{{\rm t}h})$ will be used
afterwards: by means of a weight $\exp( -\chi^2/2 )$, we will select models
with predictions $O^{{\rm t}h}$, close to the experimental data. Note that
for simplicity we have neglected correlations among different experimental
data.

\section{Looking for model independent parameterizations}

Low energy fermion masses arise from the Yukawa interaction with the light
Higgs bosons: 
\begin{equation}
L_{Yuk}=Y_{u}^{ij}H_{u}\bar{Q}_{i}u_{j}+Y_{d}^{ij}H_{d}\bar{Q}_{i}d_{j}+{\rm %
h.c.}  \label{eq:1}
\end{equation}
In the Standard Model the Higgs bosons $H_{d}$ and $H_{u}$, are the same
particle, namely $H_{d}={\rm i}\sigma _{2}H_{u}^{\ast }$ but, in general (%
{\it e.g.} in supersymmetry), they are two distinct fields. The Yukawa
couplings $Y^{ij}$, unconstrained and incalculable in the Standard Model,
arise from more fundamental high energy Lagrangians. The Lagrangian given in
(\ref{eq:1}) is an effective Lagrangian, where only light particles appear
as physical fields: at such low energy, heavy particles can only appear as
virtual internal propagators of Feynman diagrams of the full theory. For
example a process with $n$ light particles $\psi $ that goes into $k$ light
particles $\phi $ through the virtual exchange of \ one heavy field $F$ \
(with mass $m$) through the interaction $g_{1}\psi ^{n}F+g_{2}\phi ^{k}F$
can be described by one single operator $g_{1}g_{2}\;\psi ^{n}\phi
^{k}/M^{2} $ containing only the light particles. Similarly, the Yukawa
interactions in equation (\ref{eq:1}) can arise from higher dimensional
operators $H_{u}\bar{Q}_{i}u_{j}\phi ^{k}/M^{k}$ \ where the field $\phi $\
acquires a {\it vev }$ \bar \phi${\it \ }and thus $Y_{u}^{ij}=\bar{\phi}%
^{k}/M^{k}$. In general the heavy particles content could be very complex, \
but if \ the original Lagrangian is invariant under some (flavor) symmetry,
the above mentioned \ low energy operators must obey the same symmetry.
Having in mind this physical mechanism, we will show in the next section how
some specific symmetries can lead to the observed fermion mass hierarchies.

\subsection{Abelian symmetries}

In the following we will introduce a parameterization for the quark mass
matrices which is inspired by models with Abelian symmetries. To exemplify
the physical origin of such parameterization we discuss the simplest case of
one additional $U(1)_X$ at the (very high) scale $\Lambda_{new}$; after we
will comment on the more general case.

This extra $U(1)_{X}$ \ must be broken, and thus we need at least one Higgs
boson $h$ with $U(1)_{X}$-charge $q\neq 0$. This field $h$ must also be a $%
SU(2)_{{\rm weak}}\times U(1)$ singlet, otherwise the electroweak group
would be broken at too high energies. In the quark sector we call $%
q_{iL}^{u} $, $q_{iR}^{u}$, $q_{iL}^{d}$ and $q_{iR}^{d}$ the $U(1)_{X}$%
-charges of, respectively, the up and down type quarks (left and right) of
the $i$-th family. Also the light Higgs bosons (responsible for the
electroweak symmetry breaking) can have non-zero charges $q^{H_{u,d}}$ with
respect\footnote{%
As already mentioned, in the Standard Model, we have $q^{H_{u}}=-q^{H_{d}}$
because there is only one Higgs field.} to the new $U(1)_{X}$ symmetry. At
the very high energy scale, the Lagrangian includes some higher order
effective operators\footnote{%
Here we have not included operators containing both $h$ and charged
conjugated fields $h^{\ast }$. This is mandatory in supersymmetric theories,
in non-supersymmetric models these terms will be considered as subleading.} 
\begin{equation}
{\it L}_{{\rm Yukawa}}={\rm \ }g_{u}^{ij}H_{u}\bar{Q}%
_{i}u_{j}h^{n_{ij}^{u}}+g_{d}^{ij}H_{d}\bar{Q}_{i}d_{j}h^{n_{ij}^{d}}+{\rm %
h.c.}  \label{eq:2}
\end{equation}
The constants $g^{ij}$ would be calculable if masses and interactions of the
full theory were known, otherwise we can only estimate their order of
magnitude by simple dimensional analysis 
\begin{equation}
g^{ij}_{u,d}\simeq \frac{1}{M^{n_{ij}^{u,d}}}  \label{eq:5}
\end{equation}
where $M$ is the mass scale of heavy particles and the exponent ${n_{ij}}$
is such that the mass dimension of the Lagrangian (\ref{eq:2}) is equal to
four (the field $h$ has mass dimension 1). In equation (\ref{eq:2}) we have
listed all possible operators that are responsible for the generation of
quark masses. Since the field $h$ is a $SU(3)_{{\rm color}}\times SU(2)_{%
{\rm weak}}\times U(1)$ singlet, these operators are invariant with respect
to the ordinary gauge transformations for any value of ($%
n_{ij}^{u},~n_{ij}^{d}$), but if we also require the $U(1)_{X}$ invariance
only few values of $n_{ij}^{u},~n_{ij}^{d}$ are allowed; namely those which
satisfy the equations 
\begin{eqnarray}
q^{H_{u}}-q_{iL}^{u}+q_{jR}^{u}+n_{ij}^{u} q &=&0  \label{eq:4} \\
q^{H_{d}}-q_{iL}^{d}+q_{jR}^{d}+n_{ij}^{d} q &=&0  \label{eq:4b}
\end{eqnarray}
in order to guarantee that the sum of field charges is zero for each Yukawa
interaction in (\ref{eq:2}). 
For simplicity, in the following, we will consider only the case in which
all solutions of the equations (\ref{eq:4},\ref{eq:4b}) correspond to
non-negative integers\footnote{%
If $n_{ij}<0$ one can build the invariant operator $\bar{Q}q {(\bar h)}^{-
n_{ij}}$ where $\bar h$ is the charge conjugated of $h$. However, in
supersymmetry this operator is not allowed for the holomorphy of the
superpotential; thus, the corresponding entry $(i,j)$ in the quark mass
matrix would be zero.}, $n_{ij}>0$. Solving equations (\ref{eq:4},\ref{eq:4b}%
), one can build an invariant operator $\bar{Q}q h^{n_{ij}}$. In all other
cases no invariant exists.

The breaking of the $U(1)_{X}$ symmetry is expected to occur at very high
energy, where the field $h\ $ acquires a vacuum expectation value ({\it %
vev), i.e.} $\langle h\rangle =v_{X}\gg M_{W}$ ; while it is at the weak
scale that light Higgs(es) \ take the {\it vev'}s $\langle H_{u}\rangle
=v_{u}$ , $\langle H_{d}\rangle =v_{d}$ respectively. \ As a consequence,
the Lagrangian (\ref{eq:2}) \ and the estimate (\ref{eq:5}) yield the
following masses for the up sector 
\begin{equation}
m_{ij}^{u}=g_{u}^{ij}\;v_{u}\left( v_{X}\right) ^{\frac{%
q_{iL}^{u}-q_{jR}^{u}+q_{H}^{u}}{q}}\simeq v_{u}\;\epsilon ^{\frac{%
q_{iL}^{u}-q_{jR}^{u}+q_{H}^{u}}{q}};{\rm \ \ \ with \ }\epsilon =\frac{v_{X}%
}{M}  \label{eq:6}
\end{equation}
and a similar expression for the down sector. We remind that $q_{iL}$ and $%
q_{jR}$ are respectively the left-handed and right-handed fermions $U(1)_{X}$%
-charges, which are in general different ($q_{iL}^{u,d}-q_{iR}^{u,d}\neq 0$%
). If $\epsilon =v_{X}/M\ll 1$, then equation (\ref{eq:6}) implies a
definite hierarchical structure of the entries in the mass matrix, whose
orders of magnitude are proportional to the difference between the quark $%
U(1)_{X}$-charges $q_{iL}^{u}-q_{jR}^{u}$. When all charges are known,
applying (\ref{eq:6}), one predicts the order of magnitude of all masses and
mixings. The converse statement is not true: even if all mass matrix entries
are known, only some linear combination of the $q_{iL,R}^{u}$ can be derived
from (\ref{eq:6}). It is preferable to replace the $q_{iL,R}^{u}$ by the
following parameters 
\begin{equation}
\matrix{u&=&\epsilon^{({q^u_{1L}-q^u_{1R}+q^u_{3R}-q^u_{3L}})/{q}} ,
&c=&\epsilon^{({q^u_{2L}-q^u_{2R}+q^u_{3R}-q^u_{3L}})/{q}}\cr \cr
t_u&=&\frac{q^u_{3L}-q^u_{1R}}{q^u_{1L}-q^u_{1R}+q^u_{3R}-q^u_{3L}}
&k_u=&\frac{q^u_{3L}-q^u_{2R}}{q^u_{2L}-q^u_{2R}+q^u_{3R}-q^u_{3L}} }.
\label{eq:11}
\end{equation}
With these definitions, the up quark mass matrix becomes 
\begin{equation}
v_{u}\epsilon ^{\frac{q_{3L}^{u}-q_{3R}^{u}+q_{H}^{u}}{q}}\pmatrix{ a_1^{u}
\, u & &a_2^{u} \, u^{1-t_u} c^{k_u}& &a_3^{u} \, u^{1-t_u}\cr a_4^{u} \,
u^{t_u} c^{1-k_u} & & a_5^{u} \, c && a_6^{u} \, c^{1-k_u} \cr a_7^{u} \,
u^{t_u}& &a_8^{u} \, c^{k_u}& &a_9^{u} }.  \label{eq:Abeliansym1}
\end{equation}
A similar matrix can be written for the down quarks, with four additional
parameters defined from the down quark analogue of equation (\ref{eq:11})

\begin{equation}
v_{d}\epsilon ^{\frac{q_{3L}^{d}-q_{3R}^{d}+q_{H}^{d}}{q}}\pmatrix{ a_1^d \,
d & &a_2^d \, d^{1-t_d} s^{k_d}& &a_3^d \, d^{1-t_d}\cr a_4^d \, d^{t_d}
s^{1-k_d} & & a_5^d \, s && a_6^d \, s^{1-k_d} \cr a_7^d \, d^{t_d}& &a_8^d \,
s^{k_d}& &a_9^d }  \label{eq:Abeliansym2}
\end{equation}
For each matrix entry in (\ref{eq:Abeliansym1},\ref{eq:Abeliansym2}), a
complex coefficient $a_{i}^{u,d}$ with $|a_{i}^{u,d}|\simeq 1$ has been
included to take into account theoretical uncertainties coming from the very
heavy particle states (see equation (\ref{eq:6}), and the discussion above),
which are intrinsic of any low energy effective description. The matrix in
equation (\ref{eq:Abeliansym1}) has three different eigenvalues; it is easy
to prove that their order of magnitudes are equal to the three diagonal
elements in (\ref{eq:Abeliansym1}). The variables $u$ and $c$ are
consequently constrained to be of the order of the ratios $m_{up}/m_{top}$
and $m_{charm}/m_{top}$ respectively. The off-diagonal elements in equation (%
\ref{eq:Abeliansym1}) cannot be estimated by simple arguments; in fact
deriving the exponents $t_u$ and $k_u$ in equation (\ref{eq:Abeliansym1})
requires a more complex and rigorous analysis and this will be done in the
next section.

The parameterization (\ref{eq:Abeliansym1}) has been inspired by a model
with just one additional $U(1)_{X}$ symmetry and one heavy Higgs $h$; its
extent of validity is much wider but it does not cover the full set of
models based on Abelian symmetries \cite
{Ibanez:ig,Dudas:1996fe,Binetruy:1996xk,Irges:1998ax,alta}. To simplify our
analysis, we will restrict ourselves to the class of models defined by
equation (\ref{eq:Abeliansym1}), which hereafter we will simply call Abelian
symmetries. A physically interesting example is provided by the model
studied in \cite{Binetruy:1996xk,Irges:1998ax}, with three additional $%
U(1)$ factors and three new Higgses. Two of these $U(1)$ can be written as a
linear combination of the $U(1)$'s in the Cartan sub-algebra of $%
E_{6}\supset SU(5)\times U(1)_{1}\times U(1)_{2}$, and they distinguish
families. The third $U(1)_{3}$ is anomalous \cite
{Binetruy:1996xk,Irges:1998ax}. The quark matrices that immediately descend
from this symmetry choice are 
\begin{equation}
M_{u}\simeq \pmatrix{ {\lambda^8} & {\lambda^5} & {\lambda^3} \cr
{\lambda^7} & {\lambda^4} & {\lambda^2} \cr {\lambda^5} & {\lambda^2} & 1
\cr }  \label{eq:ramup}
\end{equation}
and 
\begin{equation}
M_{d}\simeq \pmatrix{ {\lambda^4} & {\lambda^3} & {\lambda^3} \cr
{\lambda^3} & {\lambda^2} & {\lambda^2} \cr {\lambda} & {1} & 1 \cr }
\label{eq:ramdwn}
\end{equation}
where $\lambda $ is an unknown parameter. One can check that these matrices
agree with the parameterizations given in (\ref{eq:Abeliansym1},\ref
{eq:Abeliansym2}), once we fix 
\begin{equation}
\begin{tabular}{cccc}
$t_{u}=5/8,$ & $k_{u}=1/2$, & $t_{d}=1/4$, & $k_{d}=0,$ \\ 
$u=\lambda ^{8},$ & $c=\lambda ^{4},$ & $d=\lambda ^{4},$ & $s=\lambda ^{2}.$%
\end{tabular}
\label{expon}
\end{equation}
Other authors \cite{alta} simply add to the usual SUSY SU(5), a U(1) flavour
symmetry that leads to a better gauge coupling unification and neutrino
phenomenology. They get 
\begin{equation}
\begin{tabular}{cccc}
$t_{u}=1/2,$ & $k_{u}=1/2$, & $t_{d}=2/5$, & $k_{d}=0,$ \\ 
$u=\lambda ^{6},$ & $c=\lambda ^{4},$ & $d=\lambda ^{5},$ & $s=\lambda ^{2}.$%
\end{tabular}
\label{expon2}
\end{equation}

\subsubsection{Fitting the mass hierarchies in the Abelian case}

{\bf The Method}\newline
The goal is to extract the values of $t_{u,d}$ and $k_{u,d}$ from the
experimental measurements. A direct fit of the data is not possible, since
the number of free parameters ($\sim 36$) in (\ref{eq:Abeliansym1},\ref
{eq:Abeliansym2}) is much larger than the number of observables, six mass
eigenvalues plus four CKM parameters.

The main obstacle comes from the coefficients $a_{i}^{u,d}$, whose phases
(and size) are not theoretically known. To cope with them, we will treat
this uncertainty as a theoretical {\it systematic error}. Namely, we have
assigned a flat probability to all the coefficients $a_{i}^{u,d}$ with 
\begin{equation}
\matrix{\frac{1}{2}<|a_i^{u,d}|<2,&0<{\mathrm{arg}}(a_i^{u,d})<2 \pi,\cr &
\cr 0<t<1, & 0<k<1. }  \label{eq:19bis}
\end{equation}
$u,c,s $ and $d$ are less than one and we randomly take them with a flat
distribution in logarithmic scale. $t_{u,d}$ and $k_{u,d}$ must satisfy the
above constraints since (by definition) we choose the entry (3,3) of the
matrices in (\ref{eq:Abeliansym1},\ref{eq:Abeliansym2}) to be the largest
one. For any random choice of the coefficients $a_{i}^{u,d}$, of the
exponents $t_{u,d},k_{u,d}$ and of the variables $u,c,d,s$ we get two
numerical matrices for, respectively, the up and the down sectors. The
diagonalization of these matrices gives  six eigenvalues, corresponding to
the physical quark masses, and two numerical unitary matrices whose
multiplication yields the CKM matrix. We have collected a large statistical
sample of events. Each one of these events can be compared with the
experimental data (see section 2) through the $\chi^{2}$: the event is
accepted with probability : 
\begin{equation}
P(O^{{\rm t}h}_i)={e^{-\frac{1}{2}\chi ^{2}(O^{{\rm t}h}_i)}}  \label{chi2}
\end{equation}
where the $\chi^2$ is defined in (\ref{eq:chisquare}). Before applying the
experimental constraints, events are homogeneously distributed in the
variables $t_{u,d},k_{u,d}$, and probability distributions are flat; but
after, applying the weight corresponding to eq. (\ref{chi2}), only points
lying in well defined regions of the space $t_{u,d},k_{u,d}$ have a good
chance to survive. \newline
Let us better clarify the reason for such not uniform distributions and the
physical interpretation of the density of points per unit area in Figures
1-2,4-7. Let us assume two different choices,\footnote{%
We will often use the word ``model'' to understand a particular and fixed
choice of the exponents $t_{u,d}$ and $k_{u,d}$.} of the exponents $t_{u,d}$
and $k_{u,d}$, that we call model 1 and model 2, lying in two different
regions (1 and 2) of the $k_{u,d}$ vs $t_{u,d}$ plane (fig.1); the Monte Carlo
generates two samples of up/down matrices, through equations (\ref
{eq:Abeliansym1},\ref{eq:Abeliansym2}). Only a fraction $p_{1}$ (and $p_{2}$%
) of matrices of the sample 1 (and 2) will pass the experimental constraints
(that is eq. (\ref{chi2})): $p_{1}$ ($p_{2}$) is the probability that model
1 (2) predicts masses and mixings in a range compatible with the
experiments. Then $p_{1}$ and $p_{2}$ are, respectively, proportional to the
density of points in regions 1 and 2. From them we can argued that the model
1 is $p_{1}/p_{2}$ more (or less, if $p_{1}/p_{2}<1$ ) likely than model 2.
Even if our Monte Carlo approach favours most predictive and accurate
models, we also emphasize that one should not mistake these results with
true experimental measurements. They only give us ``natural'' range of
values for the exponents $t_{u,d}$ and $k_{u,d}$. \newline
If $p_{max}$ is the density of points at the maximum, we call $R=p/p_{max}$
the ratio of the probability with respect to the value at the maximum. By
definition $R<1$. We will show  regions corresponding to different values
of R.\newline

\noindent {\bf The results}\newline
\begin{figure}[t]
\psfig{figure=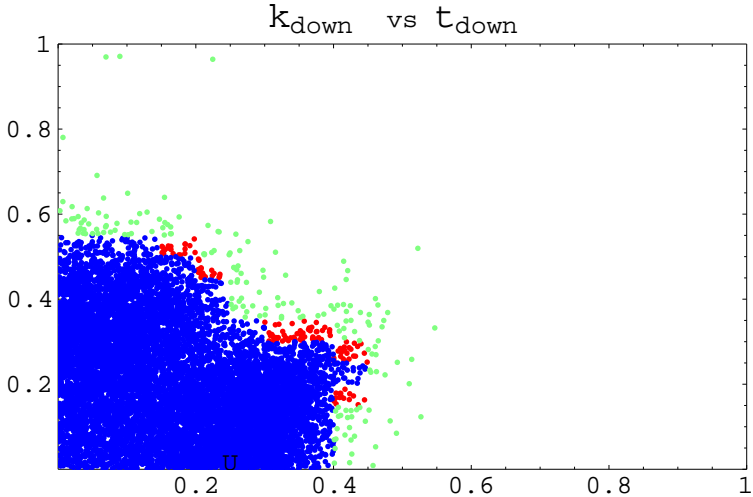,width=10cm}
\caption{Abelian Symmetries. The exponent $k_{d}$  (vertical axis) vs  $t_{d}$ 
of the parametrization given in (\ref{eq:Abeliansym2}). Points are blue (black)
(green (light gray)) in regions where $R>0.1$ ($R<0.05$) respectively}
\end{figure}

\begin{figure}[t]
\psfig{figure=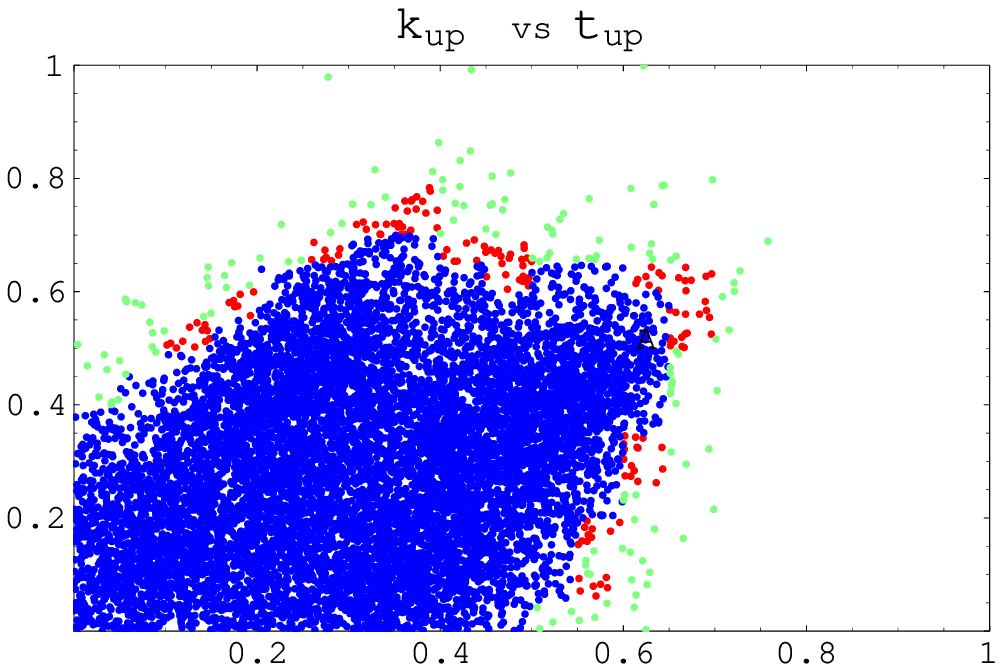,width=10cm}
\caption{Abelian Symmetries. The exponent $k_{u}$ (vertical axis) vs 
$t_{u}$ of the
parametrization given in (\ref{eq:Abeliansym1}). Points are blue (black)
(green (light gray)) in region where $R>0.1$ ($R<0.05$) respectively}
\end{figure}

Figures 1 and 2 show the density of points in the planes $k_{d}$ \ vs $t_{d}$
and \ $k_{u}$ \ vs $t_{u}.$\ Points are blue (black) in regions where $R>0.1$
and green (light gray) where $R<0.05$.\newline
A large fraction of the $t$ and $k$ space is essentially forbidden. We can
put an upper bound to all the exponents. Namely, $t_{d}\lesssim 0.55$, $%
k_{d}\lesssim 0.4$ $t_{u}\lesssim 0.7$, $k_{d}\lesssim 0.65$. This is
expected since large values for $t_{u,d}$ and $k_{u,d}$ increase the
off-diagonal elements in the CKM matrix, in contrast with the experimental
observation. The model previously mentioned \cite
{Binetruy:1996xk,Irges:1998ax}, prescribing eq. (\ref{expon}), 
lives on the border between the highly populated and the empty
regions: in the up sector (fig.2)\ a slightly higher $t_{u}$ would fall in
the not populated region. Similar conclusions, but in the down sector (fig.
1), apply for the model \cite{alta}, with exponents given in equation (\ref
{expon2}). In these specific models, a closer study of the evolution of the
Yukawa couplings (through the renormalization group equations) from the
unification scale down to the weak scale could be desirable if one wishes a
more definite conclusion. \newline
One should also note that, even if the experiments put strong bounds on the
CKM\ parameters, \ the allowed regions for $t_{u,d}$ and $k_{u,d}$ are quite
large; we have verified that a reduction on the errors of the $\bar{\rho} $ 
and $%
\bar{\eta} $ \ parameters would have a negligible impact on Figures 1 and 2. We
conclude that errors coming from the \ theoretical uncertainties (\ref
{eq:19bis}) are dominant.

For the moment, a large range of possibilities exists: even a scenario with
all exponents set to zero $t_{u,d}=k_{u,d}=0$ like in the following mass
matrices gives acceptable predictions. 
\begin{equation}
\begin{tabular}{c}
$\ \ M_{u}\propto \left( 
\begin{tabular}{ccc}
$u$ & $u$ & $u$ \\ 
$c$ & $c$ & $c$ \\ 
$1$ & $1$ & $1$%
\end{tabular}
\right) $ \\ 
$M_{d}\propto \left( 
\begin{tabular}{ccc}
$d$ & $d$ & $d$ \\ 
$s$ & $s$ & $s$ \\ 
$1$ & $1$ & $1$%
\end{tabular}
\right) $%
\end{tabular}
\label{eq:20b}
\end{equation}

\subsection{U(2) horizontal symmetry}

Another class of models is based on a $U(2)$ horizontal symmetry. This
symmetry acts on the known fermion families as follows. The light quarks
transform as doublets ${\bf 2}$ under the $U(2)$ group 
\begin{equation}
\matrix{ {\bf 2}=\pmatrix{ u_L\cr c_L\cr} & \pmatrix{ u_R^c\cr c_R^c\cr} &
\pmatrix{ d_L\cr s_L\cr} & \pmatrix{ d_R^c\cr s_R^c\cr} }.  \label{eq:12}
\end{equation}
$f_L$ are the lefthanded $SU(2)_{weak}$ doublets, while $f_R^c$ are the
charge conjugated of the righthanded $SU(2)_{weak}$ singlets. The light
Higgses (responsible for the electroweak breaking) are singlets as well as
the quarks of the third generation. The $U(2)$ (differently from $SU(2)$)
includes a $U(1)$ phase transformation: the ${\bf 2}$ and the ${\bf \bar{2}} 
$ are not equivalent representations; in particular such a $U(2)$ forbids
Yukawa interactions like \cite{ref:barbieri,Babu:1999js} 
\begin{equation}
{\rm \ } g_u H_{u}\;\overline{u}_{L}u_{R}+g_c {\rm \ }H_{u}\;\overline{c}%
_{L}c_{R}
\end{equation}
as well as all possible mass or mixing terms concerning the two lightest
generations. On the contrary the top and bottom quarks can have mass since
they are singlets under the above $U(2).$ To allow the lighter fermions
acquiring a mass, we need to break the $U(2)$ symmetry in two steps.
Firstly, the breaking of $U(2)\rightarrow U(1)$ can be induced by a $U(2)$%
-doublet $\phi _{a}$, the ${\bf 2}$, and a triplet $\Phi _{ab}$, the ${\bf 3}
$. Exploiting the $U(2)$ symmetry we can always rotate their $vev$'s in
order to obtain\footnote{%
We label the two lightest families with 1 and 2.} $\langle \phi _{2}\rangle
=v_{1}$, $\langle \Phi _{22}\rangle =v_{2}$ and $\langle \phi _{1}\rangle
=\langle \Phi _{11}\rangle =\langle \Phi _{12}\rangle =\langle \Phi
_{21}\rangle =0$ which implies the following Yukawa couplings 
\begin{equation}
{\rm \ } g H_{u}\;\overline{c}_{L}t_{R}\;\phi _{2}+g^{\prime }{\rm \ }H_{u}\;%
\overline{c}_{R}t_{L}\;\phi _{2}^{\ast }+g^{\prime \prime }{\rm \ }H_{u}\;%
\overline{c}_{L}c_{R}\;\Phi _{22}+{\rm h.c.}
\end{equation}
or in terms of the quark mass matrix 
\begin{equation}
M_{up}\propto 1/M\;\pmatrix{0 & 0 & 0\cr 0 & v_2 & v_1 \cr 0 & v_1 & M_{u} }.
\label{eq:14}
\end{equation}
An analogous matrix arises for the down sector. At lower energy also the $%
U(1)$ can be broken by a $U(2)$-singlet $A_{ab}$, antisymmetric under the
exchange of the indices $a$ and $b$. This changes the matrix (\ref{eq:14})
into 
\begin{equation}
M=\pmatrix{0 & -v_3 & 0\cr v_3 & v_2 & v_1 \cr 0 & v_1 & M_{u} }
\label{eq:15}
\end{equation}
where the scale of the $U(1)$ symmetry breaking is much smaller than the $%
U(2)$ symmetry breaking, {\it i.e.} $v_{3}\ll v_{1}\simeq v_{2}$. The zeros
in the entries 
\begin{equation}
M_{13}=M_{31}=M_{11}=0  \label{eq:16}
\end{equation}
are a generic consequence of this class of models\footnote{%
Renormalization group effect from the high energy down to the low energy can
modify this texture. A scenario with (1,3) entries slightly different from
zero has been discussed by \cite{Roberts:2001zy}.}: we will exploit the
conditions (\ref{eq:16}) to parameterize the quark mass matrices (we make no
assumption on the other entries); starting from the texture (\ref{eq:16}),
we will considerably simplify the problem of extracting all mass hierarchies
from the data; and at the same time, this will leave us with a reasonably
large and assorted selection of models. To be more concrete, after the
conditions (\ref{eq:16}) we are left with 6+6 non-zero entries that can be
parametrized by 12 free complex variables as follows (to simplify the
notation, we omit the up/down subscript in the exponents) 
\begin{equation}
M_{up}=M_{u}\pmatrix{0 & a_1^u\epsilon_1^{1-p_u} & 0\cr a_2^u
\epsilon_1^{p_u} & a_3^u \epsilon_2 & a_4^u\epsilon_2^{r_u} \cr 0 & a_5^u
\epsilon_2^{d_u} & a_6^u },  \label{eq:17}
\end{equation}
and 
\begin{equation}
M_{down}=M_{d}\pmatrix{0 & a_1^d\epsilon_3^{1-p_d} & 0\cr a_2^d
\epsilon_3^{p_d} &a_3^d \epsilon_4 &a_4^d \epsilon_4^{r_d} \cr 0 &a_5^d
\epsilon_4^{d_d} & a_6^d }.  \label{eq:18}
\end{equation}
Clearly one could choose a different reasonable parameterization with
different parameters: nevertheless the needed 12 parameters would be in one
to one correspondence with ours (shown in (\ref{eq:17},\ref{eq:18})) through
well defined equations. Thus we do not loose in generality, choosing the
above parameterization, here the only assumption\footnote{%
However this strict equivalence will become only approximately true, after
the implementation of the Monte Carlo procedure in the next section. In fact
a different parameterization would correspond to a different (and non-flat) 
{\it a-priori} distribution for the exponents $p_{u,d},r_{u,d},d_{u,d}$ (see
the beginning of the next section). This ambiguity is the necessary prize to
pay for our {\it Bayesian \ }approach.} is equation (\ref{eq:16}). The
determinant of (\ref{eq:17}) gives the product of the three eigenvalues and
thus $m_{up}m_{charm}m_{top}\sim \epsilon _{1}M_{u}^{3}.$ We also observe
that $M_{u}\sim m_{top}$, and $\epsilon _{2}\sim m_{charm}/m_{top{\rm \ }}$%
(if $r+d\gtrsim 1$), then we also get that $\epsilon _{1}\sim
m_{up}m_{charm}/m_{top}^{2}.$ A similar estimate holds for the down sector,
where $\epsilon _{3}$ and $\epsilon _{4}$ \ are related to the down quark
masses. The exponents $p_{u,d},r_{u,d},d_{u,d}$ need a more complex analysis
that will be done in the next section.

The coefficients $a_{i}^{u,d}$ take into account the intrinsic theoretical
uncertainties due to unknown fundamental parameters of the high energy
physics. They are expected to be of order one : $|a_{i}^{u,d}|\simeq 1$ with 
$0<{\rm \arg }(a_{i}^{u,d})<2\pi $.

The $U(2)$ symmetry breaking, naively discussed above, implies definite
values for the exponents 
\begin{equation}
\begin{array}{ccc}
p_{u}=1/2, & r_{u}=1, & d_{u}=1 \\ 
p_{d}=1/2, & r_{d}=1, & d_{d}=1;
\end{array}
\label{eq:u2}
\end{equation}
however a more sophisticated theoretical study could lead to different
scenarios, as for instance if one embeds the above picture into Grand
Unified Theories. Our aim is to extract these exponents directly from the
experimentally measured quantities, in the same line as shown in section
(3.1). Before moving to this determination, we would like to show with an
example the importance of the precise determination of the $\overline{\rho}$
and $\overline{\eta}$ parameters in excluding the above solution (\ref{eq:u2}) 
for the
exponents.

\subsection{Importance of the precise determination of $\overline{\protect%
\rho}$ and $\overline{\protect\eta}$ parameters}

In this example we use the values for the exponents that correspond to the
U(2) symmetry breaking given in equation (\ref{eq:u2}). This choice implies
that \cite{ref:relat,ref:barbieri,Roberts:2001zy} : 
\begin{equation}
\begin{array}{ccc}
\frac{V_{ub}}{V_{cb}} & = & \sqrt\frac{{m_u}}{{m_c}} \\ 
\frac{V_{td}}{V_{ts}} & = & \sqrt\frac{{m_d}}{{m_s}}
\end{array}
\label{eq:masse}
\end{equation}
These relations are not exact due to the presence of the order-1
coefficients. On the other hand, as discussed in \cite{ref:barbieri}, the
larger correction is on the first relation in (\ref{eq:masse}) and is of
order 10$\%$. Equation (\ref{eq:masse}) can be used to define an allowed
region in the ($\overline{\rho}$-$\overline{\eta}$) plane as shown in Figure
(3). This region has been obtained by adding an extra 10$\%$ random
correction to the first equation in (\ref{eq:masse}) to take into account
the above-mentioned effect \cite{ref:barbieri}. In this Figure the $%
\overline{\rho}$-$\overline{\eta}$ allowed region is compared with the one
selected by the measurements of $\left|\varepsilon_K \right|$, $\left | {\rm %
V}_{ub} / {\rm V}_{cb} \right |$, $\Delta m_d$ and from the limit on $\Delta
m_s$. The two selected regions are far to be compatible (to be more
quantitative see next paragraph). The significant improvement in the
determination of the $\overline{\rho}$ and $\overline{\eta}$ parameters,
obtained recently, allows to draw clear conclusions on the U(2) horizontal
symmetries. As it was anticipated in \cite{ref:barbieri} the high limit on
the $B_s$-$\bar{B_s}$ mixing parameter, $\Delta m_s$, disfavours the {\it %
naive} U(2) symmetry breaking.

\begin{figure}[htb]
{%
\epsfig{figure=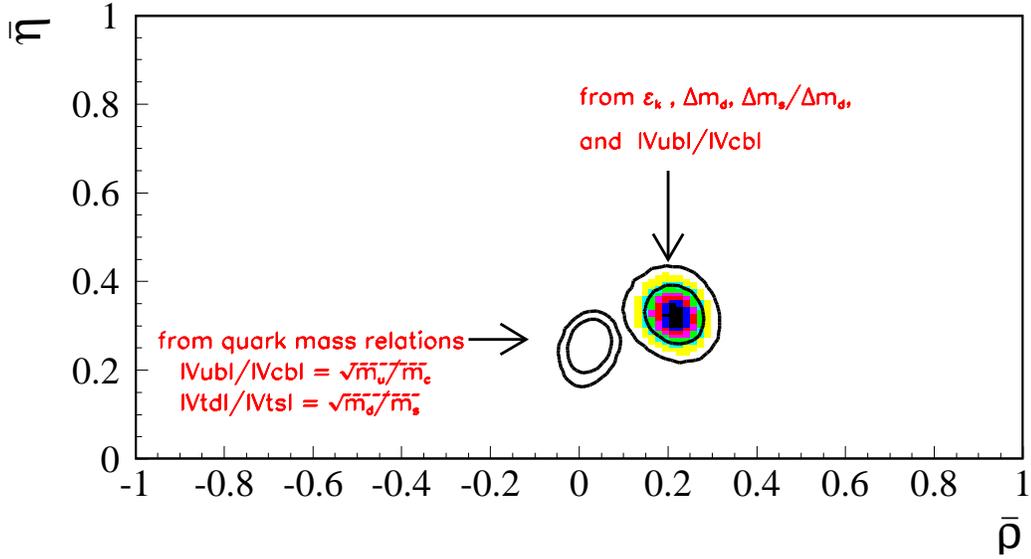,bbllx=10pt,bburx=480pt,bblly=10pt,
bbury=280pt,height=8cm}}
\caption{The allowed region for $\overline{\protect\rho}$ and $\overline{%
\protect\eta}$ (the contours at 68 $\%$ and 95 $\%$ probability are shown). 
{\it Coloured contours} : regions obtained using the constraints given by
the measurements of $\left | V_{ub} \right |/\left | V_{cb} \right |$, $%
\left|\protect\varepsilon_K \right|$, $\Delta m_d$ and $\Delta m_s$. {\it %
Empty contours} : regions obtained using the relation given in equation (\ref
{eq:masse}). }
\end{figure}

In the next section we extract the exponents of a generic U(2) horizontal
symmetry directly from the experimentally measured quantities.

\subsection{Fitting the mass hierarchies in the horizontal U(2) models}

\begin{figure}[t]
\psfig{figure=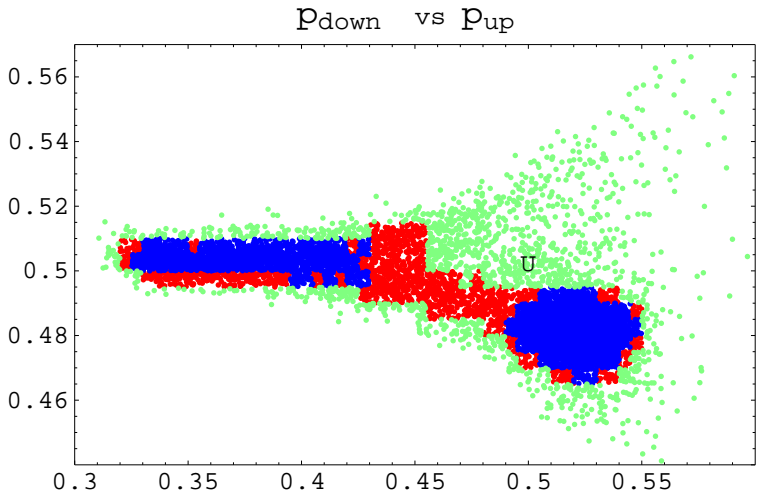,width=10cm}
\caption{ U(2) Horizontal Symmetries. The exponent $p_{d}$ (vertical axis)
 vs $p_{u}$. The
letter `` U'' marks the {\it naive} $U(2)$ prediction discussed in the text.}
\end{figure}
\begin{figure}[t]
\psfig{figure=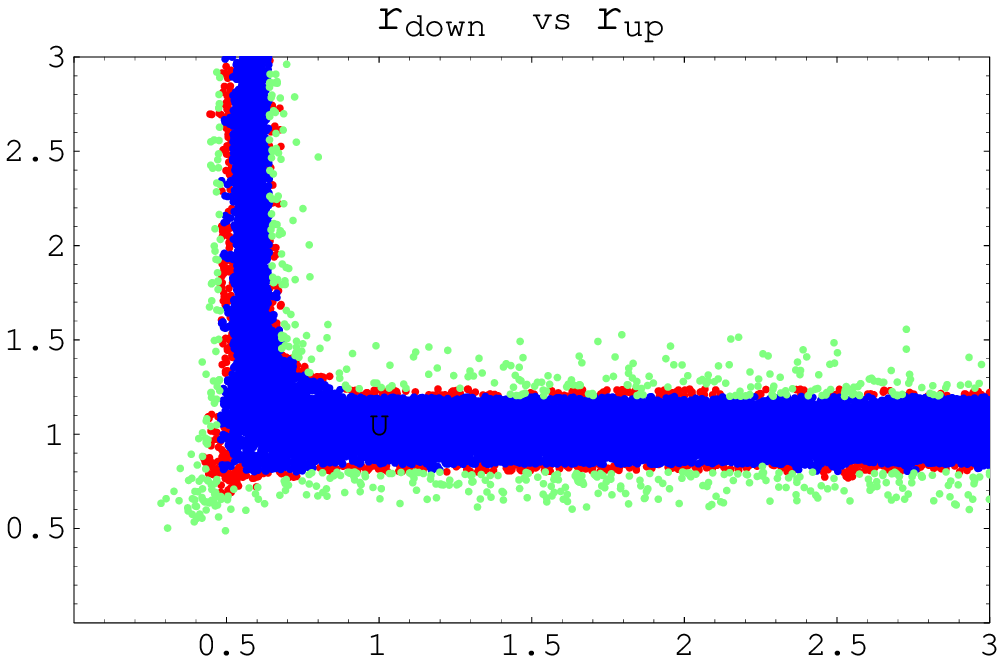,width=10cm}
\caption{U(2) Horizontal Symmetries. The exponent $r_{d}$ (vertical axis) 
vs $r_{u}$ }
\end{figure}
\begin{figure}[t]
\psfig{figure=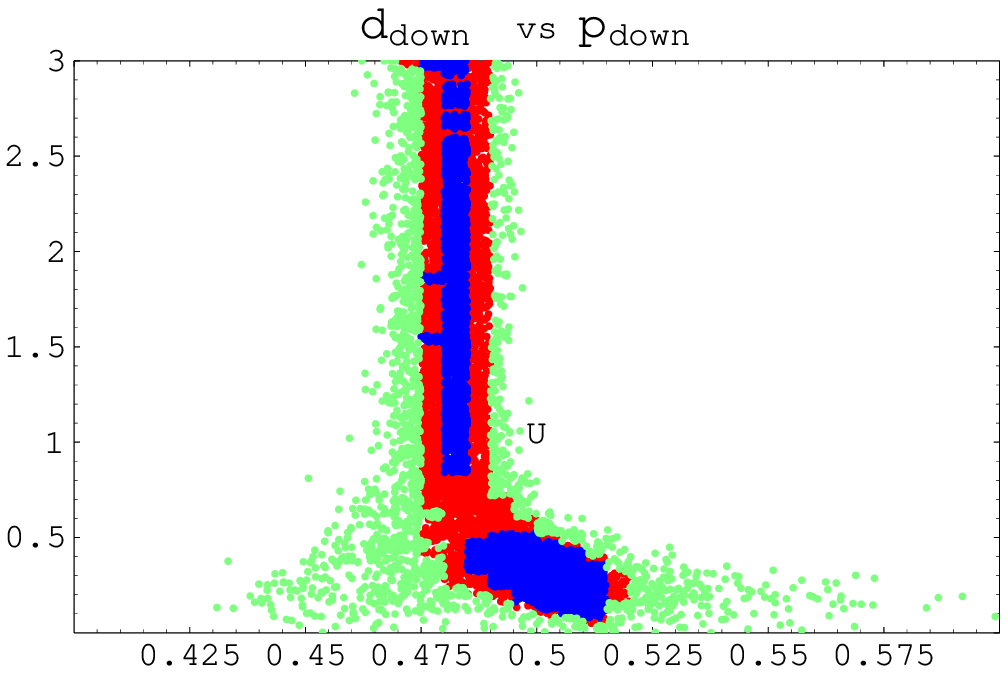,width=10cm}
\caption{U(2) Horizontal Symmetries. The exponent $d_{d}$ (vertical axis) 
vs $p_{d}$. The
letter `` U'' marks the {\it naive} $U(2)$ prediction discussed in the text.}
\end{figure}
\begin{figure}[t]
\psfig{figure=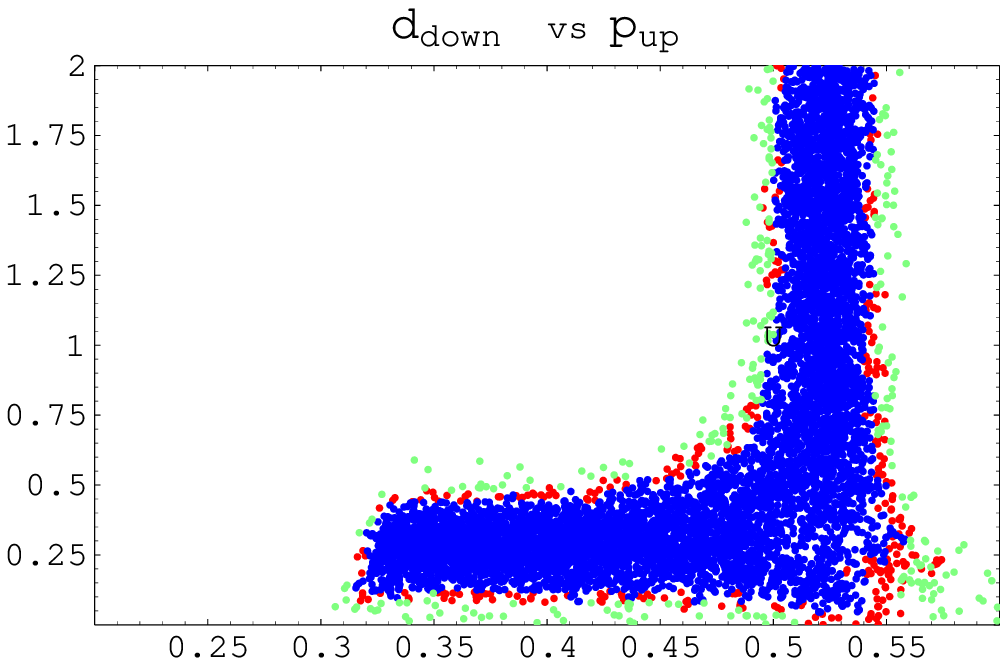,width=10cm}
\caption{U(2) Horizontal Symmetries. The exponent $d_{d}$ (vertical axis) 
vs $p_{u}$ }
\end{figure}
The $a^{u,d}\simeq 1$ coefficients in eq. (\ref{eq:17},\ref{eq:18})
(differently from the Abelian case) are redundant: for example, if $%
|a_{4}^{u}|\neq 1$, one can reabsorb the norm $|a_{4}^{u}|$ in eq. (\ref
{eq:17}) into a redefinition of $r_{u}\rightarrow r_{u}+\log
|a_{4}^{u}|/\log \epsilon _{2}$ of the exponent $r_{u}$; thus one could set $%
|a_{i}^{u,d}|=1$ without loosing in generality.

Namely, we have assigned a flat probability to all the coefficients $%
a_{i}^{u,d}$ with 
\begin{equation}
\matrix{|a_i^{u,d}|=1, &0<{\mathrm{arg}}(a_i^{u,d})<2 \pi,\cr & \cr
0<t_{u,d}<1 & 0<d_{u,d},r_{u,d}<5 }  \label{eq:19}
\end{equation}
As in the Abelian case, the range of variation of $t_{u,d}$ is fixed by the
requirement that the entries (1,2) and (2,1) are smaller than the entry
(3,3) in matrices (\ref{eq:17},\ref{eq:18}). $d_{u,d}$ and $r_{u,d}$ could
go from zero to infinity, but we have checked (see after) that for values
greater than five the corresponding entries were absolutely negligible. In
practice, the case $r_{u,d}>5$ (or $d_{u,d}>5$) is equivalent to $r_{u,d}=5$
(or $d_{u,d}=5$).

For any random choice of the coefficients $a_{i}^{u,d}$, of the exponents $%
p_{u,d},~r_{u,d},~d_{u,d}$ and\footnote{%
Note that $\epsilon _{1}$ and $\epsilon _{2}$ in the up sector are
independent from $\ \epsilon _{3}$ and $\ \ \epsilon _{4}$ in the down
sector. All of them are randomly chosen, with flat probability distribution
in logarithmic scale.} the $vev$ $\epsilon $'s we get six physical quark
masses and the CKM matrix. The large statistical sample of collected events
has been compared with the experimental data (see section 2) and a weight
defined for each event, corresponding to the probability distribution 
(\ref{chi2}) . We have found that the overall probability
distribution is maximal in regions shown in Table 3, even if it is still
high and acceptable in a larger domain.

\vskip 0.5cm

\begin{center}
\begin{tabular}{cc}
\begin{tabular}{|c||c|}
\hline
\begin{tabular}{c|c}
$p_{u}$ & $\sim 0.33$ \\ \hline
$d_{u}$ & $\gtrsim 0.3$ \\ \hline
$r_{u}$ & $\sim 0.5$%
\end{tabular}
& 
\begin{tabular}{c|c}
$\ \ p_{d}$ & $\sim 0.5$ \\ \hline
$d_{d}$ & $\sim 0.15\div 0.35$ \\ \hline
$r_{d}$ & $> 1$%
\end{tabular}
\\ \hline
\end{tabular}
& 
\begin{tabular}{|c||c|}
\hline
\begin{tabular}{c|c}
$p_{u}$ & $\sim 0.33$ \\ \hline
$d_{u}$ & $\gtrsim 0$ \\ \hline
$r_{u}$ & $\gtrsim 1.$%
\end{tabular}
& 
\begin{tabular}{c|c}
$\ \ p_{d}$ & $\sim 0.5$ \\ \hline
$d_{d}$ & $\sim 0.15\div 0.35$ \\ \hline
$r_{d}$ & $\sim 1$%
\end{tabular}
\\ \hline
\end{tabular}
\\ 
&  \\ 
\begin{tabular}{|c||c|}
\hline
\begin{tabular}{c|c}
$p_{u}$ & $\sim 0.52$ \\ \hline
$d_{u}$ & $\gtrsim 0.5$ \\ \hline
$r_{u}$ & $\sim 0.5$%
\end{tabular}
& 
\begin{tabular}{c|c}
$\ \ p_{d}$ & $\sim 0.48$ \\ \hline
$d_{d}$ & $0.6\div 1.2$ \\ \hline
$r_{d}$ & $> 1$%
\end{tabular}
\\ \hline
\end{tabular}
& 
\begin{tabular}{|c||c|}
\hline
\begin{tabular}{c|c}
$p_{u}$ & $\sim 0.52$ \\ \hline
$d_{u}$ & $\gtrsim 0.3$ \\ \hline
$r_{u}$ & $\gtrsim 1$%
\end{tabular}
& 
\begin{tabular}{c|c}
$\ \ p_{d}$ & $\sim 0.48$ \\ \hline
$d_{d}$ & $0.6 \div 0.9 $ \\ \hline
$r_{d}$ & $0.6 \div 0.9$%
\end{tabular}
\\ \hline
\end{tabular}
\\ 
& 
\end{tabular}

Table 3: Four regions in the space of the exponents $%
p_{u,d},~r_{u,d},~d_{u,d}$ that maximize the probability distribution.
\end{center}

The two solutions on top of Table 3, prefer asymmetric quark mass matrices: $%
p_{u}\sim 0.33$ implies that the entry $(1,2)$ (in the up sector) is roughly
the square of the entry $(2,1)$; also $d_{d}<r_{d}$ gives an entry (3,2)
much larger than the (2,3) (in the down sector).

The remaining two solutions (bottom row of Table 3) appears more symmetric:
a slight asymmetry is due to the exponents $p_{u}=0.52$ and $p_{d}=0.48$.
But these are very close to 1/2, thus giving approximately symmetric
matrices.

The six exponents show a significant statistical correlation; unfortunately
we can only project the sample of events on two dimensional plots, thus in
Figures 4-7, the number of points per unit area tells us the probability
distribution as function of \ each pair \ of the variables $%
p_{u,d},~r_{u,d},~d_{u,d}$ (while the remaining four variables are
integrated out); In the Figures , points are blue (black) in regions where $%
R>0.1$, red (gray) in regions where $0.05<R<0.1$, and green (light gray)
where $R<0.05$. In Figure 4 we see the points distribution in the plane $%
p_{d}$ vs $p_{u} $. Note that the allowed region is small, in particular $%
p_{d}$ is very well determined and $R>0.05$ only if $0.46\lesssim
p_{d}\lesssim 0.51 $. The exponent in the down sector $p_{d}$ is better
measured than the up one, and its value is close to $1/2$ which yields the
well known relation 
\begin{equation}
\sqrt{m_{d}/m_{s}}\simeq \sin \theta _{c}.  \label{eq:21}
\end{equation}
The parameterization for the up and the down sector are completely
equivalent but experimental data prefer to constrain the down sector, while $%
p_{u}$ can vary from 0.32 to 0.55. With U we have marked the special case of
a {\it naive } $U(2)$ symmetry as in (\ref{eq:u2}).\newline
In Figure 5 $r_{d}$ and $\ r_{u}$ are strongly correlated: either $r_{d}\sim
1$ \ or $r_{u}\sim 0.5.$ \ \ In fact if $r_{d}\sim 1$ \ and $r_{u}\gg 0.5$
one would have $V_{cb}\sim m_{s}/m_{b}$; instead $r_{u}\sim 0.5$ \ and \ $%
r_{d}\gg 1$ yield $V_{cb}\sim \sqrt{m_{c}/m_{t}}.$ Both are possible and at
least one of them must hold. In Figure 6 ($d_{d}$ vs $p_{d}$)\ \ one clearly
sees that the point U\ (referring to the model (\ref{eq:u2})) is outside the
populated region. One can overcome this problem decreasing $d_{d}$ towards
the region with maximal density $d_{d}\lesssim 0.5.$ \ Note, however, that $%
r_{d}\gtrsim 1$ (fig.5) and $d_{d}\lesssim 0.5$ imply \ \cite{Roberts:2001zy}
an asymmetric down quark matrix \ (\ref{eq:18}). In addition, decreasing $%
d_{d}$ \ one also pushes $p_{d}$ very close to 0.5, and (from Figure 4) this
solution slightly prefers $p_{u}\lesssim 0.4,$ that is the region with
maximal density, as also shown in Table 3. Also Figure 7 put in evidence a
similar behavior: $\ p_{u}$ can easily range from 0.32 to 0.55, for low
values of $d_{d}$; instead points with $d_{d}\gtrsim 0.5$ live in a much
more restricted area \ with $0.5\lesssim p_{u}\lesssim 0.55.$ \ \ Before
concluding this section it is worthwhile to note that our approach can be
used to test if a zero entry is favoured or disfavoured by data. For
example, in Figure 6 \ one can see that for $d_{d}\gtrsim 1$ \ the shape of
the point distribution no longer depends from $d_{d}$ and remains constant \
even for $d_{d}$ going to infinity. This clearly indicates that when the
entry $M_{32}=\epsilon _{4}^{d}$ \ becomes negligible ({\it i.e.} zero) few
acceptable models still survive, thus $M^{{\rm d}own}_{32}=0$ is not
excluded by data (provided that $p_{d}$ is slightly reduced).\newline

\section{ Conclusions}

The aim of particle physics is to describe all existing phenomena with the
smallest number of free parameters. In fact one usually expects that a
deeper understanding of the foundations of the theory automatically implies
a reduction of the number of free parameters. For the same reason, but in \
a different context, we await for grand unification of forces, which
replaces three gauge couplings with just one gauge coupling of the embedding
(simple) group.

Differently from the electroweak boson masses, in the fermion sector there
are several free parameters. Understanding their origin in terms of new
broken symmetries is difficult because the number of measurements is much
less than the number of such parameters. The consequent ambiguity in
extracting the eighteen complex entries of the quark mass matrices makes
very difficult to disentangle the true underlying (broken) symmetry of the
fermion sector.{\bf \ }An approximate analytical approach, \ like those
based on estimates of the observables in terms of powers of the Cabibbo
angle, can become inaccurate when several physical observables are involved:
if \ a fine tuning of\ the coefficients{\bf \ }$a$'s of order 1/3 is
required \ to get the right $V_{cb}$ ,\ one can safely conclude that the
model is acceptable; but \ if this tuning has to be repeated for \ each
observable, \ the model becomes unacceptable. We have concluded that a
global analysis based on a Monte Carlo \ procedure with model independent
parameterizations can give us more clear and trustable conclusions. We have
shown how such an approach can be successfully implemented, achieving more
information with less theoretical prejudice.{\bf \ } The progress brought by
this {\it Bayesian } approach is the fact that we can put some {\it %
confidence levels} to some important parameters which otherwise would be
unknown.We can test how precise are the determinations of the different
exponents, and this makes clear to model builders which one of the
theoretical ingredients is essential and which is not.

In case of Abelian symmetries we have found that the exponents $t$ and $k $
are poorly constrained. Still, an upper bound can be given to the four
exponents: $t_{{\rm d}}\lesssim 0.55 $, $k_{{\rm d}}\lesssim 0.4$ $t_{{\rm u}%
}\lesssim 0.7 $, $k_{{\rm d}}\lesssim 0.65$. On the contrary, no lower bound
arises from our analysis. The models explicitly discussed in the text lie in
the highly populated region, even if near its boundary.

In the non Abelian scenario, namely when we (only) assume the (1,1), (1,3)
and (3,1) entries to be zero, much stronger constraints can be set; also
strong correlations between exponents of the down and up sectors arise. The
naive $U(2)$ predictions appear ruled out. Instead, two alternative
scenarios are suggested by current data: one where $d_{{\rm d}}$ is
preferably around $0.25$. This value contradicts the naive $U(2)$ prediction 
$d_{{\rm d}}=1$. At the same time $r_{{\rm d}}\gtrsim 1,$ and \ that
requires an asymmetric texture $M_{32}^{{\rm d}}\gg M_{23}^{{\rm d}}$. This
agrees with the analysis \cite{Roberts:2001zy} . Such large $M_{32}^{{\rm d}}
$ , in GUT\ theories \cite{Anderson:1993fe}, \ implies a large (2,3) entry
in the neutrino Dirac mass which, in turn, favours a large mixing in the
atmospheric neutrinos \cite{alta,Roberts:2001zy,Berezhiani:1998vn}. 

A second scenario allows us to keep the same hierachies as in the naive $U(2)
$ , but with \  a slight (left-right) asymmetry in the (1,2)\ (2,1) entries
(both in the up sector and in the down sector). \ Namely $|M_{12}^{{\rm u}%
}|\sim 2|M_{21}^{{\rm u}}|$ and  $|M_{12}^{{\rm d}}|\sim 0.7|M_{21}^{{\rm d}%
}|$ .

\section*{Acknowledgments}

We would like to thank P. Binetruy for very helpful discussion. During this 
work,  we have also  benefited from very  interesting  discussions 
 at the GDR workshop.

\end{document}